\documentclass[prx,aps,superscriptaddress,secnumarabic,amsmath,amssymb,twocolumn,notitlepage,longbibliography]{revtex4-2}
\usepackage[colorlinks=true,urlcolor=blue]{hyperref}
\usepackage[normalem]{ulem}
\usepackage[utf8]{inputenc}
\usepackage[english]{babel}
\usepackage[T1]{fontenc}
\usepackage{mathtools}
\usepackage{mathrsfs}
\usepackage{amsfonts}
\usepackage{bbm}
\usepackage{amsthm}
\usepackage{xcolor}
\usepackage{bbold}
\usepackage{bm}
\usepackage{soul}

\renewcommand{\thefigure}{\arabic{figure}}

\definecolor{green}{rgb}{0,0.5,0}

\newcommand{\kB}{\kappa_{\rm B}}
\newcommand{\kG}{\kappa_{\rm G}}
\newcommand{\kF}{\kappa_{\rm F}}

\newcommand{\Sf}{\bm{\sigma}^{(\phi)}}
\newcommand{\Sh}{\bm{\sigma}^{({\rm h})}}
\newcommand{\Sp}{\bm{\sigma}^{({\rm p})}_{n}}
\newcommand{\Sa}{\bm{\sigma}^{({\rm a})}}

\DeclareMathOperator\tr{tr}
\DeclareMathOperator\artanh{artanh}

\begin{document}

\title{Tuneable defect-curvature coupling and topological transitions in active shells}
\author{Ludwig A. Hoffmann}
\thanks{These authors contributed equally to this work.}
\affiliation{Instituut-Lorentz, Universiteit Leiden, P.O. Box 9506, 2300 RA Leiden, The Netherlands}
\author{Livio Nicola Carenza}
\thanks{These authors contributed equally to this work.}
\affiliation{Instituut-Lorentz, Universiteit Leiden, P.O. Box 9506, 2300 RA Leiden, The Netherlands}
\author{Luca Giomi}
\email{giomi@lorentz.leidenuniv.nl}
\affiliation{Instituut-Lorentz, Universiteit Leiden, P.O. Box 9506, 2300 RA Leiden, The Netherlands}
\date{\today}

\begin{abstract}
Recent experimental observations have suggested that topological defects can facilitate the creation of sharp features in developing embryos. Whereas these observations echo established knowledge about the interplay between geometry and topology in two-dimensional {\em passive} liquid crystals, the role of activity has mostly remained unexplored. In this article we focus on deformable shells consisting of either polar or nematic active liquid crystals and demonstrate that activity renders the mechanical coupling between defects and curvature much more involved and versatile than previously thought. Using a combination of linear stability analysis and three-dimensional computational fluid dynamics, we demonstrate that such a coupling can in fact be tuned, depending on the type of liquid crystal order, the specific structure of the defect (i.e. asters or vortices) and the nature of the active forces. In polar systems, this can drive a spectacular transition from spherical to toroidal topology, in the presence of large extensile activity. Our analysis strengthens the idea that defects could serve as {\em topological morphogens} and provides a number of predictions that could be tested in {\em in vitro} studies, for instance in the context of {\em organoids}.
\end{abstract}

\maketitle

\section{Introduction}
The development of features at the early stage of embryogenesis is one of the most spectacular phenomena in developmental biology and tissue biophysics. During this process cells collectively flow over length scales orders of magnitude larger than the typical cellular size (see e.g. Refs. \cite{Streichan2018,Munster2019}), with neither external guidance nor central control mechanism, eventually giving rise to specific and reproducible morphological features~\cite{Streichan2018,Munster2019,Kiehart2017,Behrndt2012,Jain2019,MaroudasSacks2021a,Livshits2021,MaroudasSacks2021b,Aigouy2010,Etournay2015,Lecuit2007,Guillot2013,MaroudasSacks2021b}. Whereas the activity of each subunit is finely regulated by the cell's mechanosensing machinery, how this is integrated on the scale of hundreds of cells to achieve a robust and efficient morphogenetic strategy challenges our current understanding of self-organization in living matter~\cite{Saw2017,Collado2022b}. 

One of the most interesting and far-reaching concepts in this respect revolves around the hypothesis that topological defects could serve as organizing centers for morphogenetic events~\cite{Keber2014,Saw2017,MaroudasSacks2021a,Hoffmann2022}. Topological defects are isolated singularities in the orientation field of arbitrary anisotropic fluids where orientational order is locally suppressed, thereby enhancing the affinity for biological and mechanical activity. Recent {\em in vitro} experiments have suggested, for instance, that certain kinds of nematic defects, known as $+1/2$ disclinations, could govern cell apoptosis and extrusion in epithelial monolayers~\cite{Saw2017,Loewe2020,Monfared2021}. With respect to the development of features, defects have been likened to {\em topological morphogens}~\cite{Keber2014,Guillamat2022,AlIzzi2021,Metselaar2019,Pearce2020,MaroudasSacks2021a,Ruske2021,Vafa2021,AlIzzi2022,Hoffmann2022}, where the mechanical coupling between the topological charge of the defect and the local curvature~\cite{Lidmar2003,Bowick2009} conspire with the modulation of the lateral pressure caused by the collective cellular flow toward rendering the substrate unstable to buckling~\cite{Hoffmann2022}.

\begin{figure}[b!]
\centering
\includegraphics[width=\columnwidth]{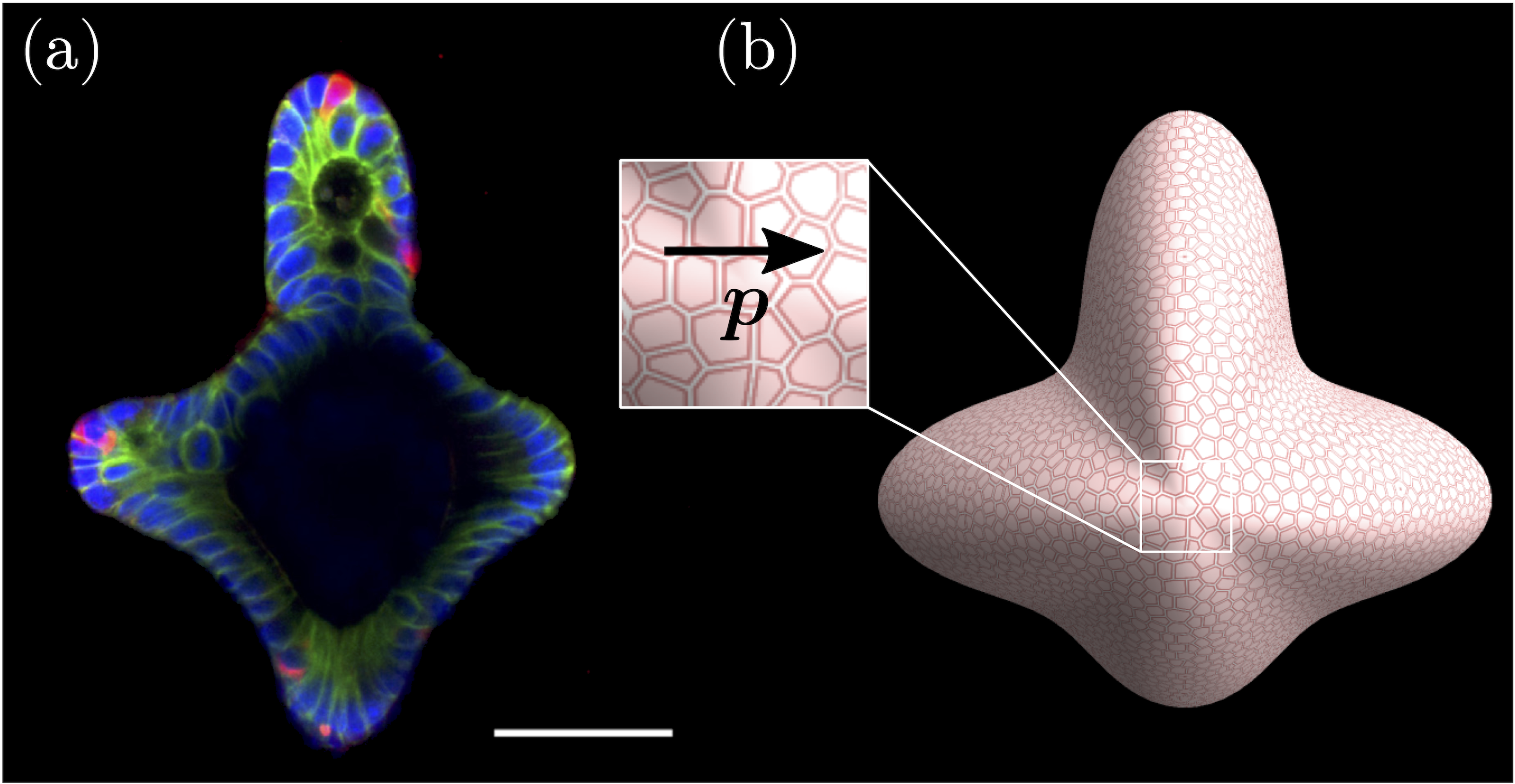}
\caption{\label{fig:fig1}(a) Example of an intestinal organoid; blue and green dyes denote the nuclei and E-cadherin respectively. Scale bar: 30$\mu$m. Adapted from Ref.~\cite{Gjorevski2022}. (b) Illustration of our continuum approach based on the hydrodynamic of active gels. The vector $\bm{p}$ indicate the local cellular orientation.}	
\end{figure}

In this article we focus on active shells $-$ i.e. thin films of active liquid crystal with spherical topology $-$ and demonstrate that the mechanical coupling between defects and curvature is in fact much more involved and versatile than previously thought. Using a combination of linear stability analysis and three-dimensional computational fluid dynamics, we show that activity can drastically affect the coupling between curvature and topological defects. In particular, while in passive media positively charged defects always~\cite{Lidmar2003,Bowick2009} elicit the formation of positively curved features, such as bumps, corners and cusps, in active liquid crystals this coupling can be tuned: depending on the magnitude and the type of activity (i.e. contractile or extensile), defects drive the appearance of features with either like- or opposite-sign Gaussian curvature, as well as instabilities with no counterpart in passive liquid crystals.
Moreover, while many living systems exhibit nematic symmetry, the minimal  excitations of a nematic liquid crystal (namely semi-integer defects) are topologically protected from escaping in the third dimension~\cite{degennes1993,chaikin1995principles}. Conversely, systems with polar symmetry, possibly arising from directed motion in cellular systems, can naturally escape in the third dimension to release the stress accumulated in proximity of topological defects. Motivated by these observations, in the following we will consider active liquid crystal with both polar and nematic symmetry. We will show that in the case of polar systems, asters and vortices, respectively, cause the sharpening and the flattening of the substrate and that, for large extensile activity, the latter mechanism can drive a remarkable transition from a spherical to a toroidal topology. By contrast, in the case of active nematics, extensile activity drives the emergence of periodic deformations and protrusions, whereas contractile activity promotes a global spindle-like shape. A natural testing ground of our predictions is found in the context of {\em organoids}~\cite{Lancaster2014,Clevers2016} $-$ i.e. that is {\em in vitro} cell aggregates with the small scale anatomy of real organs $-$ most often consisting of a cell monolayer enclosing a lumen (Fig.~\ref{fig:fig1}a). 

\begin{figure*}[t!]
    \centering
    \includegraphics[width=0.98\textwidth]{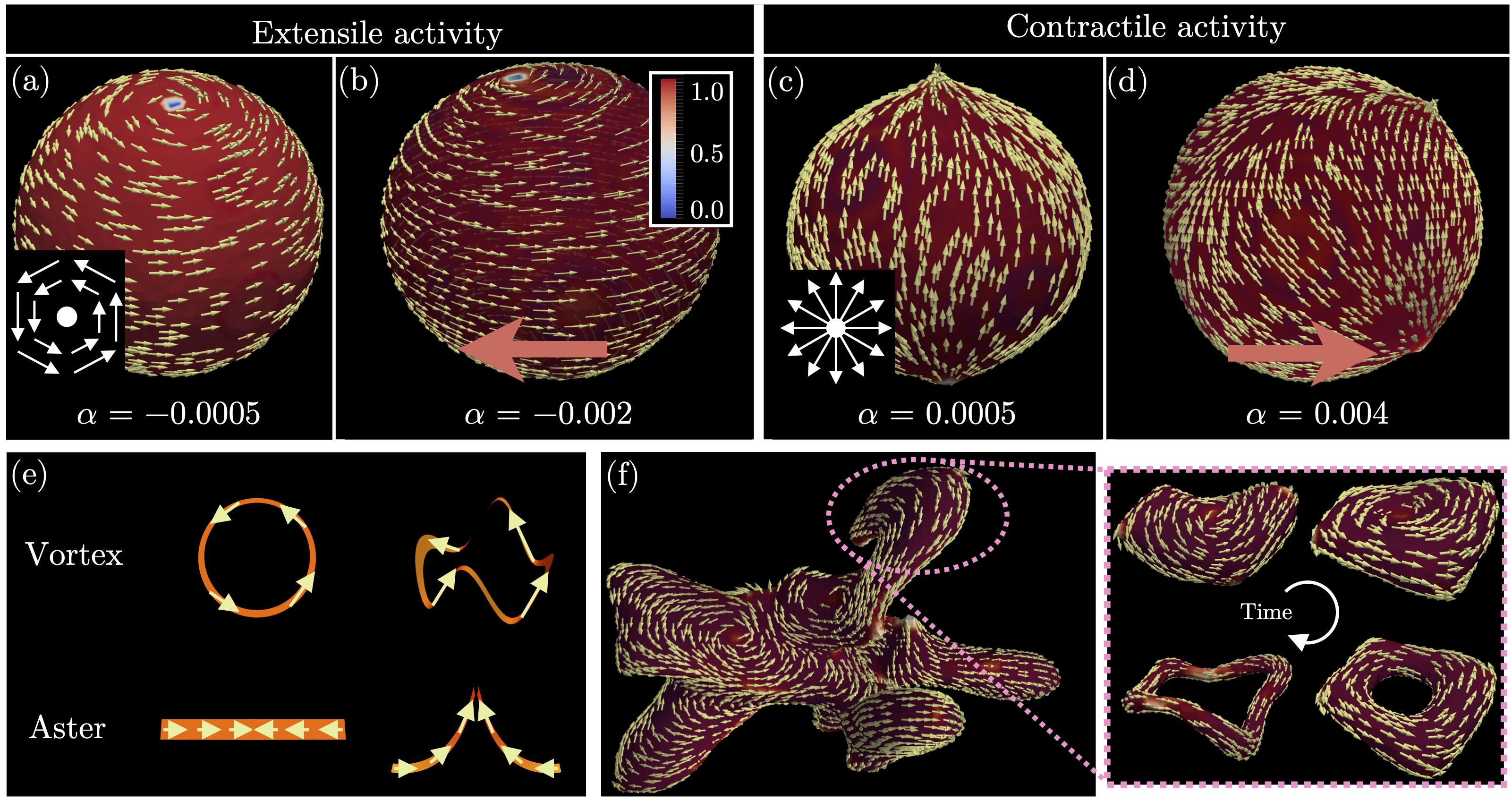}
    \caption{\label{fig:fig2}\textbf{Active polar LC on elastic shell.} (a-d)  Typical configuration of active shells and polar LC arrangements for extensile activity, panels (a-b), and contractile activity, panels (c-d).
The yellow vectors denote the local direction $\bm{p}$ polarization field whose magnitude $\Psi_1$ is used to color the shell surface according to the color bar in (b). (a) For small extensile activity ($\alpha = -0.0005$) defects (blue) with the local geometry of a vortex (sketched in inset) are located at the poles of an undeformed sphere and create an azimuthal flow. In (b), for larger extensile activity ($\alpha = -0.002$), the sphere flattens and the defects move away from the poles to increase the bending instability of the director field. This symmetry breaking results in the shell moving in the direction of the red arrow. (c) On the other hand, in the case of small contractile activity ($\alpha = 0.0005$) we observe buckling near the aster-shaped defects (aster sketched in inset). (d) Similarly as for extensile activity, for larger contractile activity ($\alpha = 0.004$), the two aster-like defects move away from the poles and the shell becomes motile. (e) To phenomenologically illustrate why we observe buckling only for asters but not for vortices, we sketch the escape in the third dimension for a LC coupled to an elastic ribbon. 
The aster configuration (bottom) is able to escape with significantly less deformation of the ribbon (right column) than for the vortex (top). The yellow arrows represent the orientation of the polar liquid crystal and the orange ribbon the confining substrate. 
(f) Deformed shell in the turbulent extensile regime ($\alpha = -0.015$). The region circled in pink produces a vesicle that eventually transforms from a flattened sphere to a torus, as shown in the time-series on the right. The hole of the torus increases in time and eventually the torus rips apart into a filament. The parameters used in all simulations can be found in the Appendix~\ref{sec:numerical_methods}.}
\end{figure*}

\begin{figure}[t!]
    \centering
    \includegraphics[width=0.9\columnwidth]{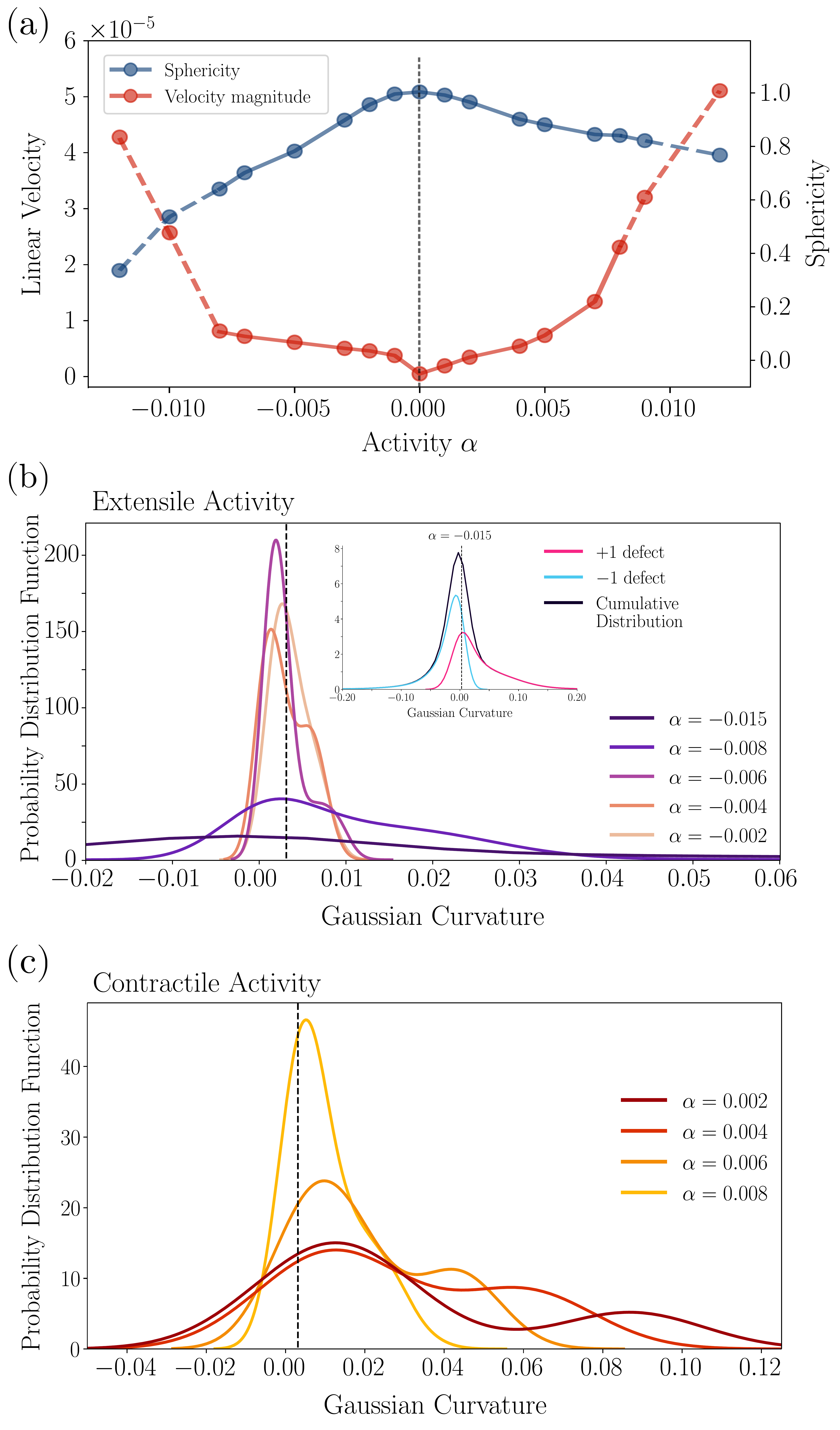}
    \caption{\label{fig:fig3} \textbf{Flattening and defect-curvature coupling for polar LC.} (a) The shell's speed and sphericity for different activities. We observe that the sphericity remains larger as contractile activity is increased, compared with the situation for increasing extensile behavior.  To better quantify our observation and justify our claim, that active defects flatten the sphere for extensile activity we consider the probability density function of the position of defects as a function of the local Gaussian curvature for different activities. (b) Probability distribution of the Gaussian curvature for extensile activity. The Gaussian curvature of the undeformed sphere of radius $R_0 = 18$ is given by $K \approx 0.003$ and indicated by the dashed line. Indeed, in this case defects (vortex) are more likely to be found in flat regions and the probability distribution function is peaked near zero Gaussian curvature.
 The inset shows the probability distribution of the Gaussian curvature for the case $\alpha=-0.015$, that is in the active turbulence regime. We present the curves for positive (magenta) and negative (cyan) defects, as well as their sum (black). On the other hand, in (c), defects in the presence of contractile activity (aster) are more likely to be found in regions where the local Gaussian curvature is greater than that for an undeformed sphere.}
\end{figure}

\section{The model}

Our model cell monolayer consists of a three-dimensional thin film of active polar or nematic liquid crystal, initially organized in the form of a spherical shell and immersed in a Newtonian solvent. The system's local configuration is described by the concentration field $\phi=\phi(\bm{r},t)$, which distinguishes the interior from the exterior of our organoid-like active shell, the incompressible velocity $\bm{v}=\bm{v}(\bm{r},t)$ (i.e. $\nabla \cdot \bm{v} = 0$), and a generic order parameter $\bm{\Psi}_{n}=\bm{\Psi}_{n}(\bm{r},t)$, with $n=1,\,2.$ For polar systems $\bm{\Psi}_{1}=\Psi_{1}\bm{p}$, with $\Psi_{1}$ a scalar order parameter and $\bm{p}$ a unit vector expressing the subunits' average direction (Fig.~\ref{fig:fig1}b). For nematics we denote the order parameter as $\bm{\Psi}_{2}$ which in case of uniaxial order~\footnote{Note that uniaxiality is assumed only to perform analytical calculations. For simulations, the fully biaxial case is considered.}
can be written as $\bm{\Psi}_{2}=\Psi_{2}(\bm{n}\bm{n}-\mathbb{1}/3)$,
with $\mathbb{1}$ the three-dimensional identity tensor, $\Psi_{2}$ a scalar order parameter and $\bm{n}$ the nematic director field. The dynamics of these material fields is assumed to be governed by the following set of hydrodynamic equations~\cite{Carenza2019}:
\begin{subequations}\label{eq:hydrodynamics_3d}
\begin{gather}
(\partial_{t}+\bm{v}\cdot\nabla)\phi = \mu\nabla^2 \left(\delta\mathcal{F}/\delta \phi \right)\;,\\[5pt]
(\partial_{t}+\bm{v}\cdot\nabla)\bm{\Psi}_{n} = \Gamma \bm{h}_n + \bm{\Omega}_{n} \;,\\
\rho(\partial_{t}+\bm{v}\cdot\nabla)\bm{v} = \nabla\cdot\left(\Sh+\Sp+\Sf+\Sa\right).
\end{gather}
\end{subequations}
Eq.~(\ref{eq:hydrodynamics_3d}a) expresses the conservation of the shell's mass, with $\mu$ a mobility coefficient and $\mathcal{F}=\int {\rm d}V\,(f_{\phi}+f_{\rm p}+f_{\rm c})$ the total free energy of the system. Here, $f_{\phi}= a (\phi/\phi_0)^2 (\phi - \phi_0)^2 + (k_{\phi}/2)|\nabla\phi|^{2}$ is the free energy density of the interface with $\phi_0$ the equilibrium value of the concentration field inside the droplet, $a$ and $k_{\phi}$ positive material parameters related to the interface thickness $\xi=\sqrt{2k_\phi/a}$ and surface tension $\gamma = \sqrt{8ak_\phi/9}$. The free energy density $f_{\rm p} \sim \kappa_F/2 |\nabla \bm{\Psi}_n|^2 + f_{\rm b}(|\nabla \phi|)$ quantifies the energetic cost associated with spatial variations of the orientation field $\bm{\Psi}_{n}$, with $\kappa_F$ the Frank constant, while the polynomial bulk free energy $f_{\rm b}$ suppresses the formation of the liquid crystal everywhere but at the droplet's interface~\footnote{We stress that, the thickness of the confining interface is small when compared to the radius of the droplet, yet non-zero. Motivated by the monolayer arrangement of experimental systems targeted by our model, in the following we will refer to the the thin active surface as \emph{model cell monolayer}.}~\cite{Carenza2022}. Its explicit form depends on the polar or nematic nature and is given in Appendix~\ref{sec:Appendix_NumericFieldEquations}. 
$f_{\rm c}$ ensures tangential anchoring of the liquid crystal across the monolayer and is given by $f_{\rm c}=W_1\,(\bm{\Psi}_{1}\cdot\nabla\phi)^{2}$ in polars and by $f_{\rm c}=W_2\,(\nabla\phi)^{\rm T}\cdot\bm{\Psi}_{2}\cdot(\nabla\phi)$ in nematics, with $W_n$ a positive constant.
 The anchoring term $f_{\rm c}$ in the free energy acts as a soft constraint for the liquid crystal field to be tangential to the surface defined by the $\phi$-field interface. Any deviation from a tangential configuration results in an increase in energy resulting in an effective force acting towards aligning the liquid crystal tangentially to the surface.

The dynamics of the orientational order parameter, in turn, is governed by Eq.~(\ref{eq:hydrodynamics_3d}b), where the first term on the right-hand side embodies the relaxation dynamics, with $\Gamma^{-1}$ the rotational viscosity and $\bm{h}_n = -\delta \mathcal{F} / \delta \bm{\Psi}_n $ the molecular field, while the second term contains the flow alignment parameter $\lambda$ and reflects the coupling between the orientational order and flow, see Appendix~\ref{sec:Appendix_NumericFieldEquations} for the full expressions. Finally, Eq.~(\ref{eq:hydrodynamics_3d}c) implies conservation of the total momentum $\int {\rm d}V\,\rho\bm{v}$, where the total density $\rho=\rho_{\rm s}+\phi$, with $\rho_{\rm s}$ the density of the solvent, is assumed to be constant and the stress tensor has been decomposed into four contributions: $\Sh = -P\mathbb{1} + 2 \eta \bm{u}$, with $P$ the pressure, $\eta$ the shear viscosity, and $\bm{u}=[\nabla\bm{v}+(\nabla\bm{v})^{\rm T}]/2$ the strain rate tensor; $\Sf = \left(f-\phi\,\delta\mathcal{F}/\delta\phi\right)\mathbb{1}-k_{\phi}\nabla\phi\,\nabla\phi$, resulting from a deformation of the active monolayer; $\Sp$, whose expression, given in Appendix~\ref{sec:Appendix_NumericFieldEquations}, explicitly depends on the polar or nematic nature of the liquid crystal and embodies the stresses originating from a distortion of the orientation field $\bm{\Psi}_n$; and, finally, the active stress $\Sa=\alpha \bm{\Psi}_{2}$ for nematics and $\Sa=\alpha\Psi_{1}(\bm{p}\bm{p}-\mathbb{1}/3)$ for polars. Here, the constant $\alpha$ is proportional to the forces exerted by the active mesogens and models contractile or extensile stresses when positive or negative, respectively.

\section{Morphogenetic activity of asters and vortices}

We begin our analysis with polar systems, by numerically integrating Eqs.~\ref{eq:hydrodynamics_3d} (with $n=1$) by means of a hybrid lattice Boltzmann approach~\cite{carenza_review} (see Appendix~\ref{sec:numerical_methods}. The liquid crystal is randomly initialized on the shell and the configuration is evolved for different values of the activity $\alpha$. In this case the Poincar\'e-Hopf theorem~\cite{David2004,LopezLeon2011} requires the existence of at least two +1 defects which, at equilibrium (i.e. $\alpha=0$), are stationary and located at the opposite poles of the sphere, in such a way as to minimize the orientational free energy~\cite{Lubensky1992}. Furthermore, on a flexible substrate and for sufficiently low surface tension, the distortion caused by this configuration is still prohibitive and the substrate is energetically favored to focus Gaussian curvature at the poles, in such a way to compensate the angular deficit introduced by the vortices, hence morphing from spherical to spindle-like~\cite{MacKintosh1991,Park1992,Lenz2003}. Extensile activity (i.e. $\alpha<0$) affects this picture by sourcing an azimuthal flow~\footnote{The normal component of the flow field becomes relevant only in proximity of topological defects. See Fig.~\ref{fig:SIfig1}.}, which in turn reorients the orientation field, ultimately leading to the spiral configuration shown in Fig. \ref{fig:fig2}a, see also Ref.~\cite{Kruse2004}. Remarkably, increasing the magnitude of the active stress $\alpha$ drives a {\em flattening} of the substrate at the poles (Fig.~\ref{fig:fig2}b, Fig.~\ref{fig:fig3}a and Movie 1), thus an expulsion of Gaussian curvature from the regions where this is energetically favored to be maximal (Fig.~\ref{fig:fig3}b). 
In response to this deformation, the defects move away from the poles and approach each other, thereby increasing the bending of the orientation field. Moreover, the breakdown of spherical symmetry, prompts a motion of the shell in the direction of the defects (Fig.~\ref{fig:fig2}b and Fig.~\ref{fig:fig3}a). The resulting non-equilibrium steady state is similar, in principle, to the motion of two-dimensional active drops that have been studied in detail, see e.g. Refs. \cite{Tjhung2012,Giomi2014_2,Tiribocchi2015}. 

By contrast, a small contractile activity ($\alpha>0$) renders the polar liquid crystal unstable to splay deformations~\cite{Kruse2004}, thereby favoring the rearrangement of the two topologically required $+1$ defects in the form of asters (Fig.~\ref{fig:fig2}c). At intermediate activity, however, the spherical shell still undergoes the same flattening dynamics described in the extensile case, but to a lesser extent (Fig. \ref{fig:fig3}a). Additionally, the aster conformation makes it possible for the orientation field to escape toward the normal direction in the core region, so to partially ease the local distortion (Fig. \ref{fig:fig2}e), compatibly with the distribution of the Gaussian curvature at the defect position shown in Fig.~\ref{fig:fig3}c. As in extensile polar shells, the defects tend to approach each other in response to the concerted action of the active flow and the deformation of the underlying substrate, while the shell moves toward the same direction (Fig.~\ref{fig:fig2}d and Movie 3).

A geometrical survey of the Gaussian curvatures at the defect position, shown in Fig.~\ref{fig:fig3}b, surprisingly reveals that vortices are preferably found in regions of small or vanishing Gaussian curvature ($ | \alpha | \leq 0.006$ in Fig.~\ref{fig:fig3}b), whereas asters tend to locate in regions of high Gaussian curvature (Fig.~\ref{fig:fig3}c). This behavior persists even in the passive limit ($\alpha=0$) for non-spherical geometries, suggesting that it is not an active effect~\cite{Napoli2012,Napoli2012b}. However, when activity is increased the bending of the liquid crystal on the surface becomes progressively more intense, leading to non-static configurations and to the wrinkling of the droplet surface. This effect is reflected in the behavior of the Gaussian curvature at the defect location (see $\alpha=-0.008$ in Fig.~\ref{fig:fig3}b) featuring a significantly broader distribution which also extends towards negative values, compatibly with the wrinkled shape of the droplet.

At even larger extensile activity, on the other hand, the active film enters a chaotic regime\footnote{We discuss here the case for extensile activity, even if a chaotic regime is also observed for contractile systems.}. At the early stage, this process is characterized by the appearance of four ``arms'', located in proximity of the defects (Fig.~\ref{fig:fig2}f), while later, as {\em active turbulence}~\cite{Giomi2015,carenza2020_bif,Carenza2020, Alert2022,Adkins2022} builds up, additional defect pairs, also featuring asters ($+1$ defects) and saddles  ($-1$ defects) nucleate. Interestingly, the inset of Fig.~\ref{fig:fig3}b shows that in the active turbulent regime defects are found in a considerably wider range of Gaussian curvature, with the positive (negative) defects localized in regions with positive (negative) curvature.
Finally, the branching shell becomes unstable to the breakup of satellite shells (i.e. pearling), whose topology evolves in time from spherical, with two $+1$ defects at the poles, to toroidal and defect free (Fig.~\ref{fig:fig2}f and Movie 2). This topological transition originates from the fact that, in the case of small satellite shells, the aforementioned flatting at the poles eventually forces the internal leaflets to come into contact and fuse. As toroidal liquid crystals, unlike spheres, are not topologically required to have defects, this process is dynamically accessible whenever there is enough extensile activity to render the mother shell unstable to pearling. The resulting toroidal shells are, however, themselves unstable to Ostwald ripening and eventually shrink and dissolve~\cite{Berti2005,Aronovitz2012,Bonelli2019,Singh2019}. The torus configuration is thus only a transient state that is not stable on long timescales. The genus transition from a spherical to a toroidal shell is relatively independent of the exact initial geometry of the spherical shell and its elastic parameters in the sense that we find the transition for various initial configurations and parameter values if the activity exceeds a critical threshold. Only this threshold depends on, e.g., the surface tension of the shell.

\section{Flattening transition in polar shells}

To gain insight into the fascinating phenomena presented above, we have considered a reduced version of the two-dimensional limit of Eqs.~(\ref{eq:hydrodynamics_3d}b) and (\ref{eq:hydrodynamics_3d}c) (see Appendix~\ref{sec:Appendix_AnalyticEoM} for details) together with the condition
\begin{equation}
\label{eq:shape}
\Delta P - f_{\rm e}^{n} - f_{\rm p}^{n} 
= K_{ij} \left[-\Pi g^{ij} + 2 \eta u_{\parallel}^{ij} + \alpha \left(p_{\parallel}^{i}p_{\parallel}^{j}-\frac{1}{2}\,g^{ij}\right)\right]\;, 
\end{equation}
resulting, after the dimensional reduction, from force balance along the surface normal~\cite{Hoffmann2022,Salbreux2019}. Here $\Delta P$ is the Laplace pressure while $f_{\rm e}^{n}$ is the normal force originating from the restoring force due to the surface tension $\gamma$ and the bending modulus $\kappa_B$. The normal force $f_{\rm p}^{n}$ derives from the distortions of the liquid crystal tangent orientation field $\bm{p}_{\parallel}$ and is proportional to the Frank elastic constant $\kappa_F$.
On the right-hand side of Eq.~\eqref{eq:shape}, $\bm{K}$ and $\bm{g}$ are the extrinsic curvature and metric tensor, respectively~\cite{David2004}, $\Pi$ the lateral pressure acting on a two-dimensional fluid patch and $\bm{u}_{\parallel}$ the strain rate associated with the tangent velocity field $\bm{v}_{\parallel}$ (see Appendix~\ref{sec:Appendix_AnalyticEoM} for details). We stress that, in this analytical calculation, the interface is assumed strictly two-dimensional, in contrast to our previously described numerical model. As the thickness $\xi$ of the simulated shell is very small compared to the system size, however, we expect this difference to be marginal with respect to the in-plane dynamics of the director (see also Fig.~\ref{fig:SIfig1}).

\begin{figure}[t!]
    \centering
    \includegraphics[width=0.98\columnwidth]{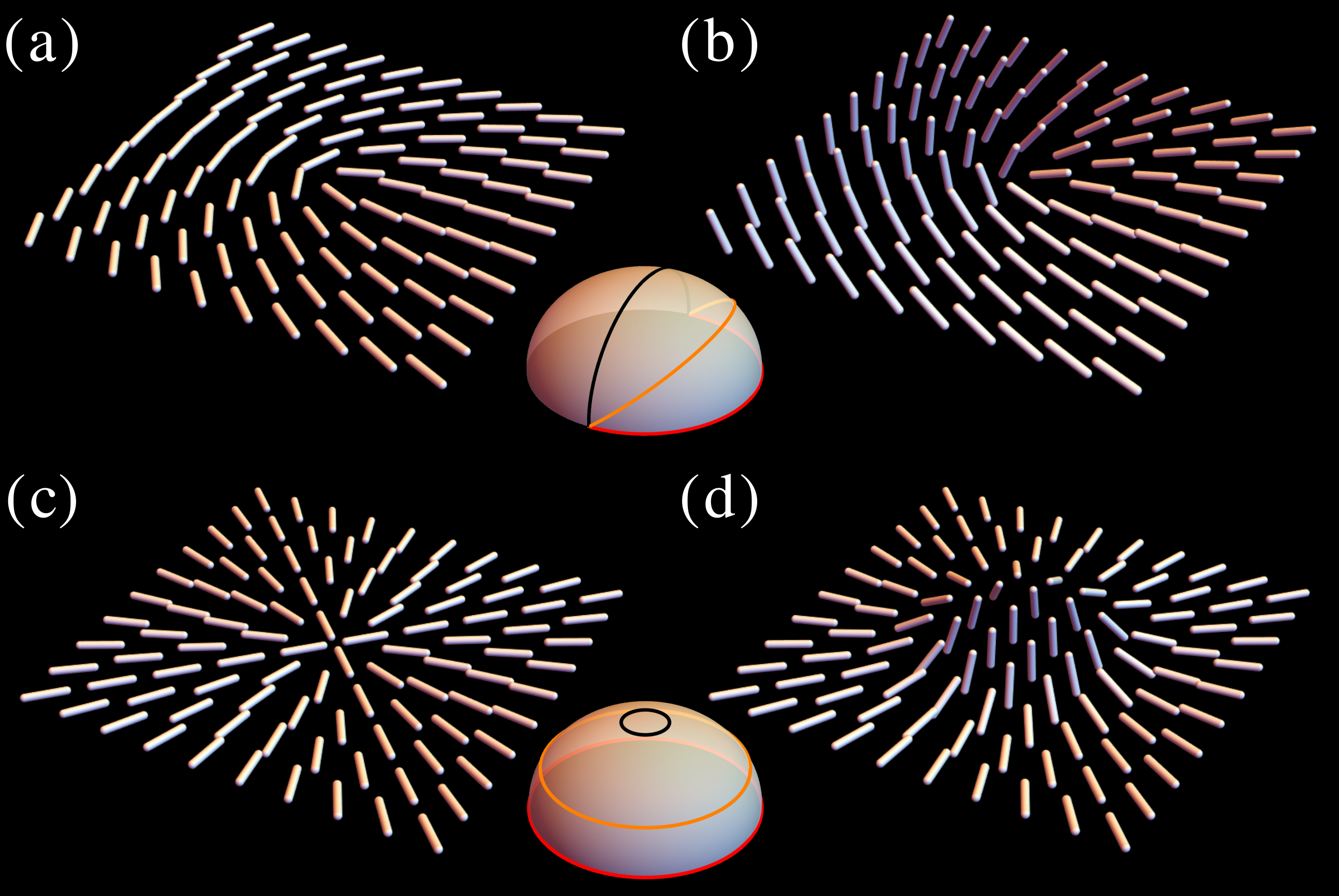}
    \caption{\label{fig:fig4}\textbf{Topological protection of nematic defects.} (a) Configuration of a $+1/2$ defect on the plane. (b) A $+1/2$ disclination can be continuously deformed into a $-1/2$ disclination upon rotating the director about an arbitrary axis on the plane of the director by an angle $\Theta=\pi$. By contrast, rotating the director by an angle $\Theta=\pi/2$ gives rise to an escaped-like configuration, but does not remove the central singularity, which is, therefore, topologically protected. The latter can be better appreciated by considering the corresponding trajectory traced by the director in one loop around the defect core and corresponding to a hemicircle connecting two antipodal points on the hemisphere (inset). These, in turn, correspond to the same orientation in the physical space because of the $D_{\infty h}$ symmetry of uniaxial nematics. In this representation, semi-integer point defects correspond to hemicircles on the equatorial plane (red curve), with their sign given by the orientation of the hemicircle with respect to that of the equator. Escaped-like configurations correspond instead to hemicircles connecting two antipodal points across the positive half-space (orange and black curve), while an isolated point identifies a uniformly oriented configuration. Whether lying on the equatorial plane or bending across the positive half-space, hemicircles cannot be contracted into a point, thus it is impossible to continuously transform a planar disclination into a defect free configuration by rotating the nematic director out of plane. (c) Configuration of a $+1$ defect on the plane. (d) Unlike for semi-integer defects, a $+1$ planar disclination can be continuously transformed into a defect free state upon rotating the director perpendicularly to the plane. In the order parameter space, this amounts to contracting a loop around the equator into a point (inset).}
\end{figure}

On a sphere of radius $R_{0}$ a stationary flowing solutions can be found, consistently with the numerical solution illustrated in Fig.~\ref{fig:fig2}a, in the form $\bm{p}_{\parallel}=(R_{0}^{-1} \cos \epsilon)\,\bm{e}_{\theta}+ (R_{0}^{-1} \sin \epsilon/\sin\theta)\,\bm{e}_{\varphi}$ and $\bm{v}_{\parallel}=\alpha/(2\eta)\,\sin 2\epsilon\,\artanh(\cos\theta)\,\bm{e}_{\varphi}$, where $\bm{e}_{\theta}$ and $\bm{e}_{\varphi}$ are tangent unit vectors in the direction of the polar angle $\theta$ and the azimuthal angle $\varphi$ respectively, while $\epsilon = 1/2\,\arccos(-1/\lambda)$ and $\Pi=\Pi_{0}=-(\alpha/\lambda)\log(\sin\theta)$. With this solution in hand, one can then consider a linear azimuthally symmetric perturbation of the sphere radius $\delta R = \delta R(\theta)$ and, after expanding this in Legendre polynomials -- i.e. $\delta R=\sum_{l=1}^{\infty}\delta R_{l}\mathcal{P}_{l}(\theta)$ -- find $\Pi=\Pi_{0}-\alpha\delta R/(\lambda R_{0})$, from which the renormalized Laplace pressure is found, from the $l=0$ mode, to be $\Delta P = \left[2 \gamma - \alpha/(\lambda \sqrt{\pi})\right]/R_{0}$. Furthermore, for a $+1$ defect located at each pole, $\delta R_{l} = 0$ for odd $l$ values, while for even $l$ values
\begin{equation}\label{eq:delta_R}
\delta R_{l} = \sqrt{(2l+1)\pi}\,\frac{2 \kappa_{\rm F} + \frac{4 \alpha R_{0}^{2}}{\lambda (l-1)(l+2)}}{l(l+1)\,\gamma R_{0} + \kB f(l)}\;,
\end{equation}
where $f(l) = l(l+1)[l^{2}(l+1)^{2}+2]/\left[(l-1)(l+2) R_{0}\right]$. The dominant mode is $l=2$ and, dropping higher modes, one finds that the curvature at the poles changes from positive to negative, thus indicating the tendency of the shell to flatten (see also Fig.~\ref{fig:SIfig2} in the Appendix), for $R_{0}$ larger
than
\begin{equation}
\label{eq:critical_radius}
R_{\rm c} = \sqrt{2|\lambda|}\, \ell_{\rm a}\;,
\end{equation}
if $\alpha/\lambda < 0$ and with $\ell_{\rm a}=\sqrt{\kF/|\alpha|}$ the so-called active length scale expressing the distance at which active and passive torques balance~\cite{Giomi2015}. Eq.~\eqref{eq:critical_radius} also holds for contractile systems, in which case, taking the formal limit $\lambda\to-1$ to recover the radial configuration of $\bm{p}_{\parallel}$ around the asters (i.e. $\epsilon=0$ and $\bm{p}_{\parallel}=\bm{e}_{\theta}$), leads again to Eq.~\eqref{eq:critical_radius}. The asymmetry between extensile and contractile shells captured by the simulations, but not by the linear stability analysis, likely results from the previously mentioned tendency of the orientation field associated with asters to escape toward the surface normal. This additional deformation mode stabilizes asters, thus rendering contractile shells less prone to flattening than extensile shells.

\begin{figure}
    \centering
    \includegraphics[width=0.85\columnwidth]{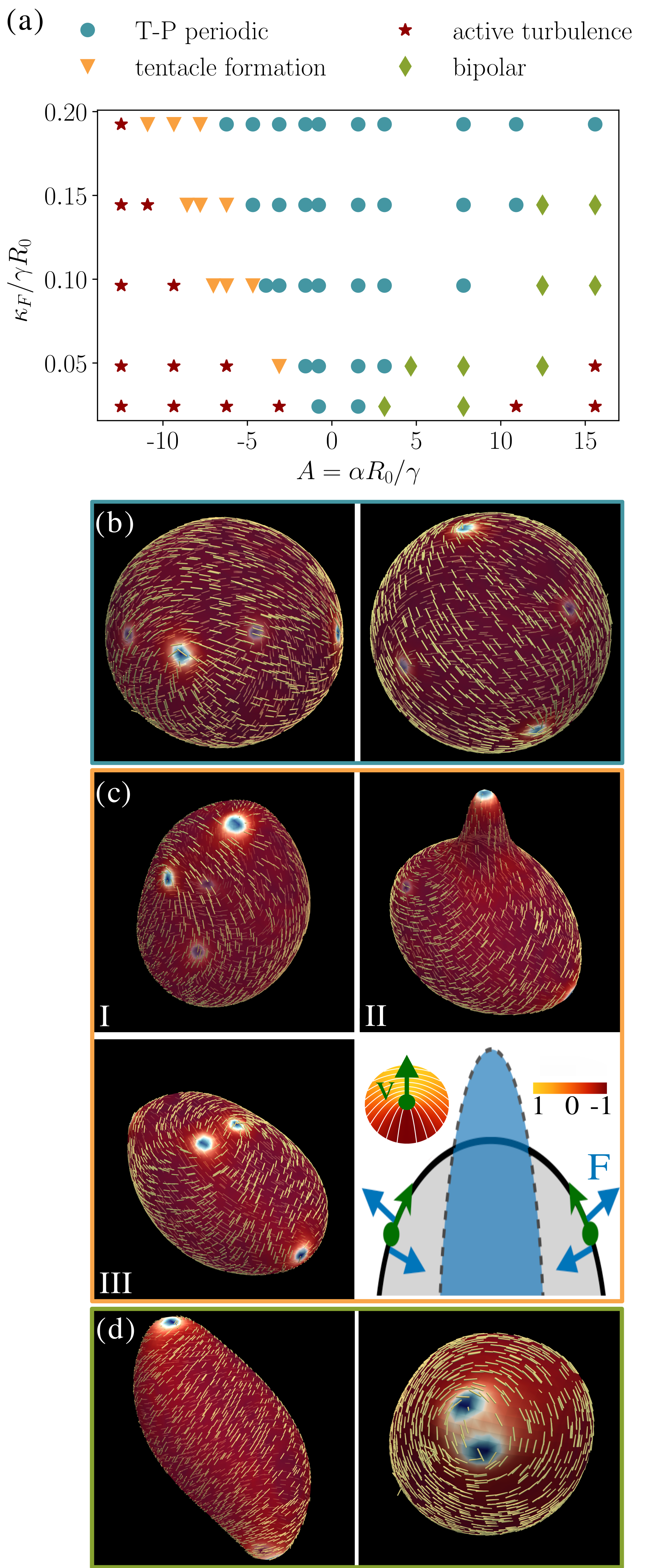}
    \caption{\label{fig:fig5}\textbf{Active nematic LC on elastic shell.} 
	We observe four distinct phases for an active nematic liquid crystal as presented in the phase diagram in (a). At small activity $+1/2$-defects move periodically on an essentially undeformed sphere (blue circles, Tetrahedral-Planar (T-P) periodic). Defects oscillate between a planar (left snapshot in panel (b)) and a tetrahedral (right snapshot in panel (b)) configuration. For larger extensile activity, defects deform the sphere periodically and, for sufficiently large activity, protrusions are formed (yellow triangles in (a)) as shown in (c) for $\alpha = -0.003$. In the main text, using the sketch in (c), it is explained how the pressure field of a $+1/2$-defect moving with velocity $\mathbf{v}$ (top-left corner) can lead to a dipolar normal force $\mathbf{F}$ (blue arrows in the sketch) that creates the protrusion. For larger contractile activity we observe a bipolar configuration (green diamonds in (a)) as shown in (d), for $\alpha = 0.002$, with two $+1/2$ defects located near each pole. For even larger activity we find a turbulent regime (red stars).}
\end{figure}

\begin{figure}
    \centering
    \includegraphics[width=0.98\columnwidth]{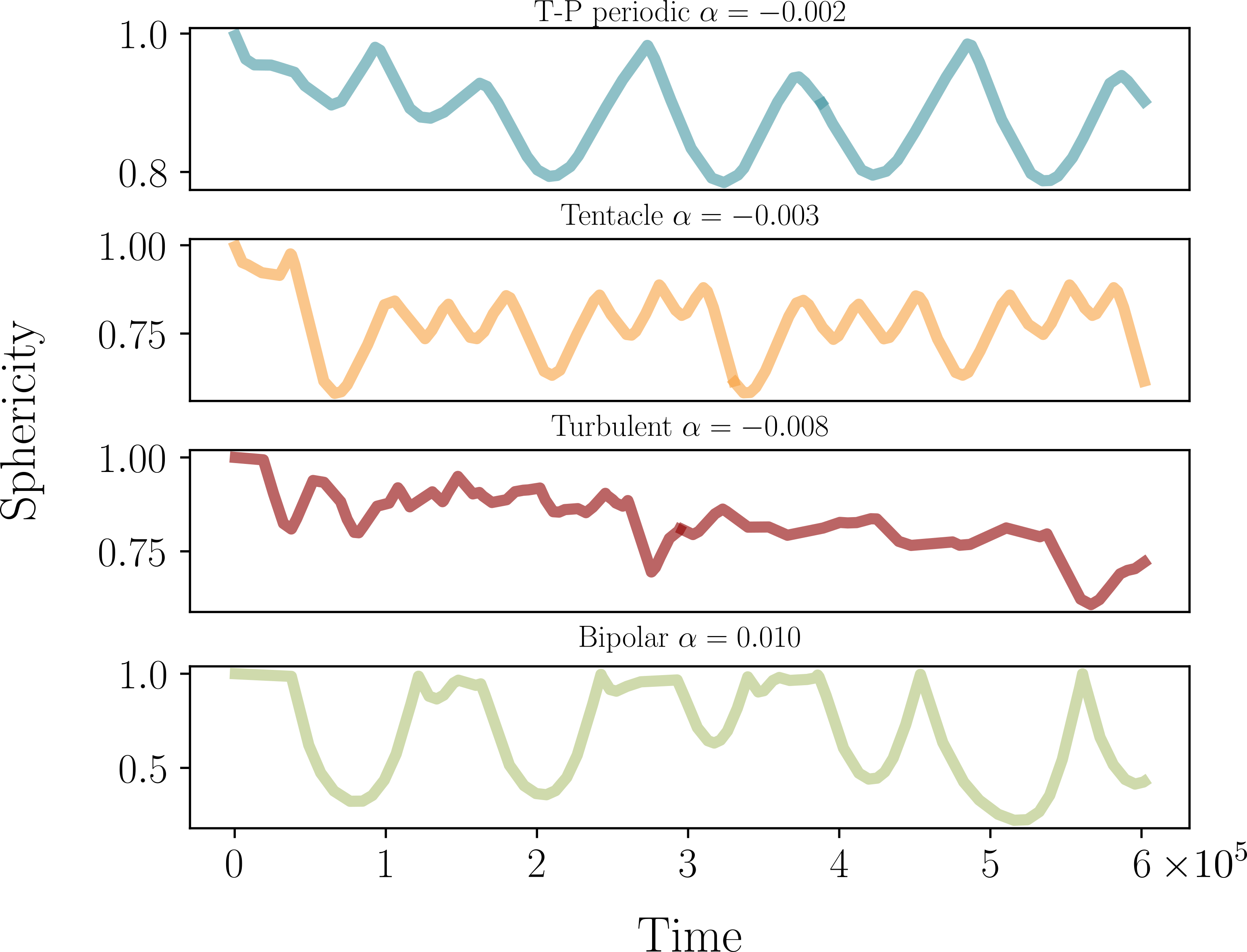}
    \caption{\label{fig:fig6}\textbf{Periodic deformation of elastic shell.} To quantify the periodic deformation of the sphere due to the active nematic defects we show the sphericity of the elastic shell as a function of time in panel for different values of activity. Each of the activity values we chose corresponds to one of the four different phases we found (see Fig. \ref{fig:fig5}). The periodicity in the sphericity is clearly visible, as is the fact that the deformation increases with increasing activity.}
\end{figure}

\section{Protrusion formation in nematic shells}

Next, we turn our attention to nematic shells, whose equilibrium configuration consists of four topologically required $+1/2$ defects~\cite{Lubensky1992,Vitelli2006,Keber2014,LopezLeon2011}.

In the following it is important to recall that, if director is allowed to escape from the tangent plane of the shell mid-surface, $+1$ defects are no longer topologically protected. This is because, by aligning the director normally to the shell mid-surface, it is always possible to {\em continuously} deform it into a defect-free configuration. This process, in turn, is contrasted by the bending elasticity of the shell, which, by reducing the local curvature of the shell, indirectly penalizes escaped configurations. The ability to escape in the third dimension also affects the topology of $\pm1/2$ defects, which can now continuously transform into each other upon rotating by $180^{\circ}$ about an axis parallel to the local tangent plane \cite{Alexander2012} (see Fig.~\ref{fig:fig4}). While in the full three-dimensional space, this feature allows, for example, the existence of neutral disclination lines~\cite{Copar2019,Duclos2020,Kralj2023} this procedure is not enough to remove the singularity, and for this reason, semi-integer defects are still topologically protected~\cite{Lubensky1992}. 

To study the effect of activity we numerically integrate Eqs.~\ref{eq:hydrodynamics_3d} for $n = 2$. At low extensile activity, when the shell preserves the initial spherical conformation, we recover the oscillatory motion initially observed in Ref.~\cite{Keber2014} in active vesicles and then thoroughly investigated using different approaches (Fig.~\ref{fig:fig5}a,b)~\cite{Zhang2016,Khoromskaia2017,Guillamat2018,hardouin2019,Henkes2018,Brown2020,Zhang2020,Nestler2021,Rajabi2021}. During one semi-period, the defects move from a tetrahedral configuration to a planar one or vice versa. For larger $|\alpha|$ values, such an oscillatory motion is inherited by the shell itself, which, again by virtue of the mechanical coupling between defects and curvature, periodically deforms from spherical to elliptical (Fig.~\ref{fig:fig5}a, Fig.~\ref{fig:fig6}, and Movie 4). Upon increasing the extensile activity further, the shape oscillations becomes more pronounced and eventually leads to the growth of protrusions (Fig.~\ref{fig:fig5}c and Movie 5). Specifically, these originate from the merging of two $+1/2$ defects into a $+1$ aster, after which the protrusion shrinks and the aster splits again into two $+1/2$ defects~\footnote{Note that even in the case of a fixed sphere the defects merge if the activity is high enough, see Ref. \cite{Henkes2018}}. Although analytically intractable, this behavior can be rationalized by considering that the lateral pressure $\Pi$ undergoes a gradient in the opposite direction compared to that of the flow velocity, thus the direction of motion of a defect. To see this note that the pressure field of an isolated $+1/2$-defect in flat space is given by $P_{\rm h} = \alpha/2 \cos \arctan(y/x)$, in a reference system centered at the defect position with the $x$ direction defining the symmetry axis of the $+1/2$ topological defects. This defect moves with velocity $\bm{v}$ in negative $x$-direction, thus in the direction of larger pressure. Locally, close to the defect core, this expression is a good approximation of the pressure field of a $+1/2$-defect even in the presence of small curvature. Since $\Delta P \sim \Pi$, when the activity is sufficiently large to overcome other restoring forces, such a gradient in the Laplace pressure  gives rise to bending moments that either sharpen the surface, if the defects move toward each other, or flatten it, if they move apart (Fig.~\ref{fig:fig5}c). In the former case, the resulting curvature increase has the secondary effect of attracting the defects, thereby enhancing the performance of the morphogenetic mechanism via a positive feedback loop. Finally, for even larger $|\alpha|$ values, the active nematic shells enters a chaotic regime, where perpetual pearling instabilities detach from the main shells several elongated, snail-like surfaces, which eventually dissolve because of Ostwald ripening (Movie 6). Similarly, at low contractile activity, the shell exhibits oscillations analogous to those discussed in the extensile case, whereas for larger $\alpha$ values, two $+1/2$ defects move towards the poles and deform the sphere into in a spindle-like shape reminiscent of a bipolar configuration and  located in the regions of high curvature (Fig.~\ref{fig:fig5}d and Movie 7).
We highlight that wide variety of dynamical states as well as defect configurations presented in the phase diagram of Fig.~\ref{fig:fig5}a is possible only because of the flow induced by activity. Indeed, regardless of the strength of surface tension and elasticity, activity can be used to select and tune a particular state.

\section{Conclusions}

In this article we have investigated the mechanical coupling between defects and curvature in active shells of polar and nematic liquid crystals as a possible morphogenetic mechanism in developing tissues. In passive materials, this coupling arises from the fact that defects introduce an orientational deficit that a like-sign Gaussian curvature can compensate, thereby reducing the system's elastic energy~\cite{Bowick2009}. As a consequence, positively charged defects, such as $+1/2$ or $+1$ disclinations in nematic and polar liquid crystals respectively, are either attracted by or able to focus positive Gaussian curvature, depending on the substrate's flexibility as well as the orientational stiffness of the fluid~\cite{Lidmar2003}. 

By contrast, in active shells the interplay between defects and curvature is much more versatile, so that the presence of $+1$ vortices at the poles of an active polar shell can either increase or decrease the local curvature, depending on whether the system is extensile or contractile. Such a correlation between regions of high (low) Gaussian curvature and $+1/2$ ($+1$) disclinations, echos a recent observation by Maroudas-Sacks {\em et al.}, who suggested that $+1/2$ defects could favor the growth of highly curved tentacles, while $+1$ defects facilitate the positioning of the gently curved mouth and foot regions in {\em Hydra}~\cite{MaroudasSacks2021a}. Furthermore, at large extensile activity, such a vortices-mediated flattening can result in a fusion of the internal leaflet of the shell, which eventually drives a transition from spherical to toroidal topology. Something related has recently be observed by Khoromskaia and Salbreux \cite{Khoromskaia2021} and in the context of elastic sheets by Pearce {\em et al.}~\cite{Pearce2020}.

To better understand the pathway leading to the formation of protrusions, we focused on active nematic shells and showed that, for large extensile activity and after two oscillatory regimes (Fig.~\ref{fig:fig6}), protrusions appears as the result of the merging of pairs of $+1/2$ disclinations into asters. Unlike previously though, this process crucially relies on the polar structure of $+1/2$ defects~\cite{Vromans2016}, which, as result of crosstalk between flow velocity and pressure along the longitudinal direction of the defect, leads to a steep and highly localized gradient in the Laplace pressure. This, in turn, gives rise to bending moments that sharpen the surface, when the two $+1/2$ defects approach each other, or flatten it when these move apart. 

Our work has a natural tie with the emerging field of organoids mechanics (see e.g. Ref.~\cite{Buchmann2021}), where topology could possibly serve as a key to deciphering the complex elongation and branching dynamics routinely observed in {\em in vitro} experiments.

\section*{Acknowledgements}
This work is supported by the Netherlands Organization for Scientific Research (NWO/OCW), as part of the Vidi scheme (L.A.H. and L.G.), and by the  European Union via the ERC-CoGgrant HexaTissue (L.N.C. and L.G.). Simulations were performed on the Dutch national e-infrastructure with the support of SURF through the Grant 2021.028 for computational time. L.A.H. thanks Ireth Garc\'ia-Aguilar for helpful discussions.

\appendix
\setcounter{figure}{0} 
\setcounter{equation}{0}     
\renewcommand{\thefigure}{A\arabic{figure}}
\renewcommand{\theequation}{A\arabic{equation}}

\vspace{1cm}

\section{\label{sec:Appendix_NumericFieldEquations}Phase-field model}

\subsection{Active polar shells}

\begin{figure*}
    \centering
    \includegraphics[width=\textwidth]{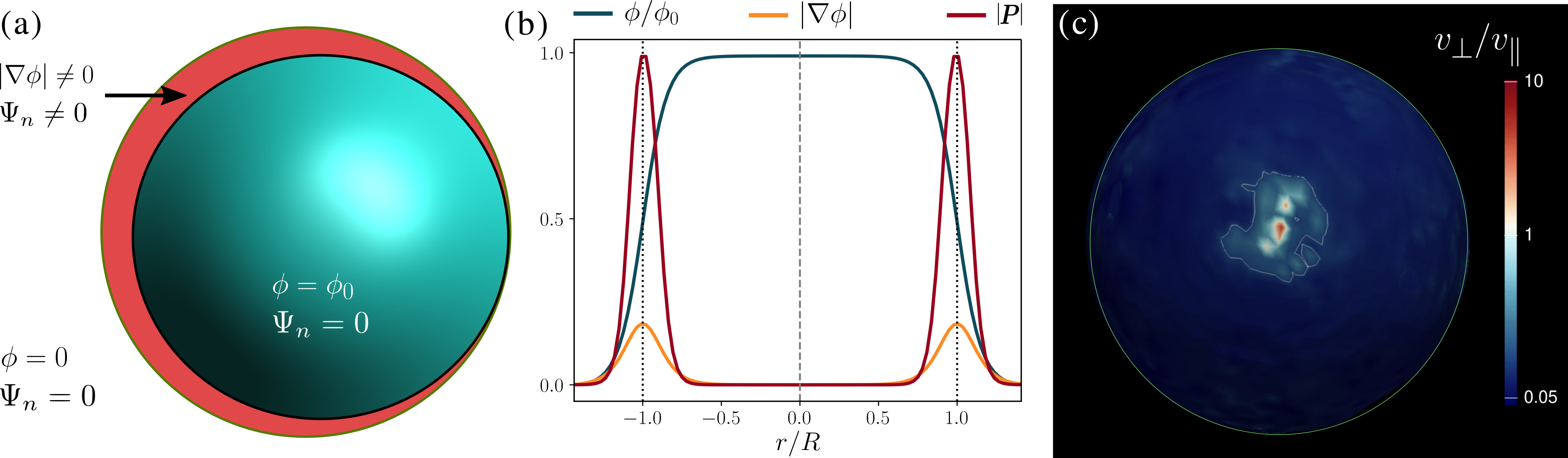}
    \caption{(a) Cross-section of an active shell, as modeled in our numerical simulations. In the interior of the droplet (plotted in green) $\phi=\phi_0$, while $\phi \approx 0$ in the exterior. Concentration gradients $\nabla \phi$ are significant only at the droplet interface (red in the picture). By construction the orientational order parameter $\bm{\Psi}_{n}$ is finite when $|\nabla \phi|$ is larger than a critical threshold (see Appendix~\ref{sec:Appendix_NumericFieldEquations}), hence the fluid is orientationally ordered and active within a thin shell which corresponds to the droplet interface and isotropic as well as passive elsewhere. (b) The radial profiles (with $r$ being the distance from the centre of the droplet) measured from simulations and expressed in lattice units of $\phi,\nabla \phi$ and the magnitude of the order parameter $|\bm{P}|$ for the polar case. Panel (c) shows the color-plot of the ratio of normal versus tangential flow $v_\perp/v_\parallel$ measured in simulations for the polar case of Fig.~\ref{fig:fig1}(b). Notice that normal flows are significant only in proximity of topological defects.}
    \label{fig:SIfig1}
\end{figure*}

To describe a polar liquid crystal we take  $\bm{\Psi}_1 = \Psi_1 \bm{p}$, with $\Psi_1$ a scalar order parameter and $\bm{p}$ the direction of the local polarization. The free energy density is then given by (see e.g. Ref.~\cite{Hoffmann2022}):
\begin{equation}
f_{\rm p} = \dfrac{\kF}{2}\,|\nabla \bm{\Psi}_1|^2 + A_0 \left( \dfrac{\psi}{2}\,|\bm{\Psi}_1|^2 + \dfrac{1}{4}\,|\bm{\Psi}_1|^4 \right)\;.
\end{equation}
The parameter $\psi$ depends on the concentration gradient $|\nabla \phi|$ and is chosen such that $\psi=-1$ if $|\nabla \phi|$ is larger than a suitable threshold $\mathcal{O}(\phi_0/\xi)$ [$0.1$ in our simulations] and zero otherwise~\cite{Hoffmann2022}, see also Fig.~\ref{fig:SIfig1}(a,b)). The bulk constant $A_0$ fixes the coherence length of the liquid crystal, i.e. $\ell_{\rm c}=\sqrt{\kF/A_0}$, which controls how sharply the order parameter drops from one to zero in proximity of a topological defect. The stress tensor associated with a distortion of the director field is
\begin{align}
\sigma^{\rm (p)}_{1} =& -\kF\nabla\bm{\Psi}_1\cdot(\nabla\bm{\Psi}_1)^{\rm T} +\frac{1}{2}\,(\bm{\Psi}_1\bm{h}_1-\bm{h}_1\bm{\Psi}_1) \nonumber \\
&-\frac{\lambda}{2}\,(\bm{\Psi}_1\bm{h}_1+\bm{h}_1\bm{\Psi}_1)\;,
\end{align}
where $\lambda$ the flow-alignment parameter and $\bm{h}_1 = - \delta \mathcal{F}/\delta \bm{\Psi}_1$ is the molecular field.
The strain-rotational derivative in the Leslie-Ericksen equation is now given by $\bm{\Omega}_1 = \lambda \bm{u} \cdot \bm{\Psi}_1 -\bm{\omega} \cdot \bm{\Psi}_1$, where $\bm{\omega}=[(\nabla\bm{v})-(\nabla\bm{v})^{\rm T}]/2$ is the vorticity tensor.

\subsection{Active nematic shells}

The order parameter to describe a nematic liquid crystal is now the $2$-ranked nematic tensor, traceless and symmetric, $\bm{\Psi}_2 = \Psi_{2}(\bm{n}\bm{n}-\mathbb{1}/3)$, with $\Psi_{2}$ the nematic order parameter. The free energy density is given by (see e.g.  Refs.~\cite{Carenza2019,Carenza2022}):
\begin{align}
f_\text{p} =& A_0 \left[ \dfrac{1}{2}  \left(1 - \dfrac{\Phi}{3} \right)|\bm{\Psi}_{2}|^2 - \dfrac{\Phi}{3} |\bm{\Psi}_{2}|^3 + \dfrac{\Phi}{4} |\bm{\Psi}_{2}|^4 \right] \nonumber \\
& + \frac{\kF}{2}\,|\nabla\bm{\Psi}_{2}|^2 .
\end{align}
The nematic liquid crystal is confined at the interface by requiring the parameter $\Phi$ to depend on the gradients of the phase field $\phi$ as follows
\begin{equation}
\Phi=\Phi_0 + \Phi_{\rm s} (\nabla \phi)^2 \;,
\end{equation}
so that $\Psi_{2}$ is non-zero only in regions where $|\nabla \phi|>\sqrt{(\Phi_{\rm c}-\Phi_{0})/\Phi_{s}}$, with $\Phi_{\rm c} = 2.7$ the critical value above which the system is in the ordered phase, and $\Phi_0$, $\Phi_{\rm s}$ free parameters~\cite{Carenza2022}. The stress tensor arising in response of a departure of the nematic order parameter tensor from its lowest free energy configuration is given by
\begin{multline}
\sigma^{\rm (p)}_{2} = -\lambda \left[\bm{h}_2 \cdot \left(\bm{\Psi}_{2} + \frac{1}{3}\,\mathbb{1}\right) + \left(\bm{\Psi}_{2} + \frac{1}{3}\, \mathbb{1}\right) \cdot \bm{h}_2 \right] \\
+ 2\lambda \left(\bm{\Psi}_{2} - \frac{1}{3}\,\mathbb{1}\right) \bm{\Psi}_{2} : \bm{h}_2 + \bm{\Psi}_{2} \cdot \bm{h}_2 - \bm{h}_2 \cdot \bm{\Psi}_{2}\;,
\end{multline}
where the molecular field $\bm{h}_2$ is defined as
\begin{equation}
\bm{h}_2 = -\frac{\delta \mathcal{F}}{\delta \bm{\Psi}_{2}} + \frac{1}{3}\,\tr\left(\frac{\delta \mathcal{F}}{\delta \bm{\Psi}_{2}}\right) \mathbb{1}\;.
\end{equation}
Finally, the strain-rotational derivative in the Leslie-Ericksen equation is given by
\begin{multline}
\bm{\Omega}_2 = (\lambda \bm{u} + \bm{\omega})\cdot\left(\bm{\Psi}_{2} + \frac{1}{3}\,\mathbb{1}\right) + \left(\bm{\Psi}_{2} + \frac{1}{3}\,\mathbb{1}\right)\cdot(\lambda \bm{u} - \bm{\omega})\\
- 2 \lambda \left(\bm{\Psi}_{2} + \frac{1}{3}\,\mathbb{1}\right) \tr(\bm{\Psi}_{2} \cdot \nabla \bm{v}) \;.
\end{multline}

\subsection{Numerical Method}
\label{sec:numerical_methods}
The dynamical equations, Eqs.~\eqref{eq:hydrodynamics_3d}, have been integrated by means of a  hybrid lattice Boltzmann (LB) method, where the hydrodynamics is solved through a \emph{predictor-corrector} LB algorithm~[34] on a $D3Q15$ lattice~\cite{succi2018}, while the dynamics of the order parameter has been treated with a finite-difference approach implementing a first-order upwind scheme and fourth-order accurate stencil for the computation of spacial derivatives.

In our LB algorithm, the evolution of the fluid is described in terms of a set of distribution functions ${f_i(\mathbf{r}_\alpha,t)}$ (with the index $i$ labelling different lattice directions, ranging from $1$ to $15$ for our D3Q15 model) defined on each lattice site $\textbf{r}_\alpha$. Their evolution follows a discretized predictor-corrector version of the Boltzmann equation in the Bhatnagar-Gross-Krook (BGK) approximation:
\begin{align}
f_i (\mathbf{r}_\alpha + \bm{\xi}_i \Delta t) - f_i (\mathbf{r}_\alpha,t) =& - \dfrac{\Delta t }{2} \left[\mathcal{C}(f_i,\mathbf{r}_\alpha, t) \right. \nonumber \\
&+ \left. \mathcal{C}(f_i^*,\mathbf{r}_\alpha+ \mathbf{\xi}_i \Delta t, t) \right].
\label{eqn:LBevolution}
\end{align} 
Here $\lbrace \bm{\xi}_i \rbrace$ is the set of discrete velocities of the $D3Q15$ lattice.
Here, $\mathcal{C}(f,\mathbf{r}_\alpha, t)=-(f_i-f_i^{eq})/\tau + F_i$ is the collisional operator in the BGK approximation, expressed in terms of the equilibrium distribution functions $f_i^{eq}$ and supplemented with an extra forcing term $F_i$ for the treatment of the anti-symmetric part of the stress tensor and $\tau$ the relaxation time of the algorithm, related to the viscosity of the fluid, as discussed in the following.
The distribution functions $f_i^*$ are first-order estimations to  $f_i (\mathbf{r}_\alpha + \bm{\xi}_i \Delta t) $ obtained by setting $f_i^* \equiv f_i$ in Eq.~\eqref{eqn:LBevolution}. 
The density and momentum of the fluid are defined in terms of the distribution functions as follows:
\begin{equation}
\sum_i f_i = \rho \qquad \sum_i f_i \bm{\xi}_i = \rho \mathbf{v}.
\label{eqn:variables_hydro}
\end{equation}
The same relations hold for the equilibrium distribution functions, thus ensuring mass and momentum conservation. 
In order to correctly reproduce the Navier-Stokes equation, the following conditions on the second moment of the equilibrium distribution functions are imposed:
\begin{equation}
\sum_i f_i \bm{\xi}_i \otimes \bm{\xi}_i = \rho \mathbf{v} \otimes \mathbf{v} -\tilde{\sigma}_s,
\label{eqn:constrain_second_moment}
\end{equation}
whilst the following condition is imposed for the force term:
\begin{equation}
\sum_i F_i = 0, \qquad \sum_i F_i \bm{\xi}_i = \mathbf{\nabla} \cdot \tilde{\sigma}_a, \qquad \sum_i F_i \bm{\xi}_i \otimes \bm{\xi}_i = 0.
\label{eqn:constraint_force}
\end{equation}
In the equations above, we respectively denoted with $\tilde{\sigma}_s$ and $\tilde{\sigma}_a$ the symmetric and anti-symmetric part of the total stress tensor, diminished of the hydrodynamic contribution $\sigma^{\rm (h)}$, which naturally arises from the continuum limit of Eq.~\eqref{eqn:LBevolution} (see Ref.~\cite{carenza_review}).
The equilibrium distribution functions are expanded up to second order in the velocities, as follows:
\begin{equation}
f_i^{eq} = A_i + B_i (\bm{\xi}_i \cdot \mathbf{v}) + C_i |\mathbf{v} |^2 + D_i (\bm{\xi}_i \cdot \mathbf{v})^2 + \tilde{G}_i : (\bm{\xi}_i \otimes \bm{\xi}_i).
\end{equation}
Here the coefficients $A_i, B_i,C_i,D_i,\tilde{G}_i$ are to be determined by imposing the conditions in Eqs.~\eqref{eqn:variables_hydro} and \eqref{eqn:constrain_second_moment}. In the continuum limit, the Navier-Stokes equation is restored  by choosing the relaxation time $\tau=3\eta/\rho$ and the speed of sound $c_s=1$ (see also Ref.~\cite{denniston2001} and~\cite{carenza_review}).

Eqs. \eqref{eq:hydrodynamics_3d} were integrated in a three-dimensional box of size $L=128$, volume $V = L^3$, and periodic boundary conditions. 
For all results presented in the main text, the radius of the shell is $R=18$. More radii, $R=15, 24, 32$ have been also simulated to check consistency of results. We report no qualitative difference with the cases presented in the main text.
The numerical code has been parallelized by means of Message Passage Interface (MPI) by dividing the computational domain in slices and
by implementing the ghost-cell method to compute derivatives on the boundary of the computational subdomains.
Runs have been performed using $64$  CPUs for at least $10^6$ lattice Boltzmann iterations (corresponding to $\sim 35d$ of CPU-time on Intel Xeon 8160 processors).
The model parameters in lattice units used for simulations are $ a=0.01,k_\phi=0.015, \phi_0=2.0,\mu=0.1, \Gamma=0.2, \eta=5/3$. For polar liquid crystals we used $A_0 =0.1, W_1=0.03, \lambda=1.1, \kappa_F=0.02$ and we varied the activity in the range $-0.015, 0.015$ as reported in the main text. For nematics we used $A_0 =0.1, \Phi_0 =2.45, \Phi_s=1.0, W_2=0.01, \lambda=1.1$. We varied the activity in the range $-0.015, 0.015$ and the Frank constant in the range $0.008, 0.04$.
The typical flow velocity measured in simulations $v\sim 3\times 10^{-2}$, in lattice units. Therefore, numerical density fluctuations $\delta \rho / \langle \rho \rangle \sim Ma^2 = (v/c_s)^2 \sim 10^{-3}$ are negligible and the fluid is effectively incompressible.

\section{Two-dimensional limit}

\subsection{\label{sec:Appendix_DifferentialGeometry}Mathematical preliminaries and notation}

A surface $\mathcal{M}$ is described by a mapping $\bm{X}=\bm{X}(s_{1},s_{2})$ and parametrized by the coordinates $(s^{1},s^{2})$ such that $\bm{e}_{i} = \partial_{i} \bm{X}$, with $\partial_{i} = \partial/\partial s^{i}$, is a local basis of tangent vectors. The metric tensor is defined as $g_{ij} = \bm{e}_{i} \cdot \bm{e}_{j}$ and the second fundamental form as $K_{ij} = -(\partial_{i}\partial_{j} \bm{X}) \cdot \bm{n}$, where $\bm{n} = \bm{e}_{1} \times \bm{e}_{2}/|\bm{e}_{1} \times \bm{e}_{2}|$ is the unit normal vector to the surface pointing outward. The Levi-Civita tensor is defined as $\epsilon_{ij} = \bm{n} \cdot (\bm{e}_{i} \times \bm{e}_{j})$. We denote the covariant derivative by $\nabla_{i}$. The normal and tangent vectors are related via the Weingarten equation $\nabla_{i}\bm{n}=K_{i}^{\;j}\bm{e}_{j}$ and the Gauss equation $\nabla_{i}\bm{e}_{j} = -K_{ij}\bm{n}$.
A general three-dimensional vector can be decomposed into a component in the tangent plane of the surface and a component perpendicular to it, $\bm{V}_\parallel = V^i \bm{e}_i$ and $\bm{V}_\perp=V_n\bm{n}$. For the mean and Gaussian curvature we use the convention $H = g^{ij}K_{ij}/2$ and $K = \det(\bm{K})$.

\subsection{\label{sec:Appendix_AnalyticEoM} Hydrodynamic equations for polar membranes}

The two-dimensional limit of Eqs.~\eqref{eq:hydrodynamics_3d} is obtained upon assuming $\Psi_{1}={\rm const}$ and treating the cell polarization as a tangent unit vector field on $\mathcal{M}$: i.e. $\bm{p}_\parallel=p^{i}_\parallel \bm{e}_{i}$, with
\begin{equation}
|\bm{p}_{\parallel}|^{2}=g_{ij}p^{i}_\parallel p^{j}_\parallel=1\;.
\end{equation}
The free energy of the system is given by
\begin{equation}\label{eq:FreeEnergy}
F = \int_\mathcal{M} {\rm d}A\,\left[ \gamma + \kB \left(H-H_0\right)^2 + \kG K+ \frac{\kF}{2}\,|\nabla\bm{p}_\parallel|^{2}\right]\;.
\end{equation}
The first three terms on the right-hand side of Eq. ~\eqref{eq:FreeEnergy}, where $\gamma$ is the surface tension, $\kB$ the bending rigidity, $H_0$ the spontaneous mean curvature and $\kG$ the Gaussian-splay modulus, comprise the Helfrich free energy~\cite{Helfrich1973}. The last term is the Frank free energy in one-elastic-constant approximation which drives the liquid crystal towards an aligned state~\cite{Chaikin1995}. Assuming the velocity field to be incompressible, a stationary configuration of the fields $\bm{p}_\parallel$ and $\bm{v}_\parallel$ is found by solving the following set of hydrodynamic equations~\cite{Giomi2015, Pearce2019, Hoffmann2022}:
\begin{gather}
\eta \left(\nabla^{2}v^{i}_{\parallel} + K v^{i}_{\parallel} \right) - \nabla^{i} \Pi+ \alpha \nabla^j p_{\parallel}^{i}p_{\parallel}^{j}  = 0 \;, \label{eq:NavierStokesSI} \\
v^k_\parallel \nabla_k p^{i}_{\parallel} = \left(g^{ij} - p^i_\parallel p^j_\parallel \right) \left(\lambda u_{\parallel jk}  \ p^k_\parallel - \omega_{\parallel jk} \ p^k_\parallel + \Gamma h_{\parallel j} \right) \label{eq:LeslieEricksenSI}.
\end{gather}
Here, Eq.~\eqref{eq:NavierStokesSI} is the covariant Stokes equation, with $\eta$ the shear viscosity, $\Pi$ the hydrodynamic pressure enforcing the incompressibility of the fluid and we have neglected the passive stress tensor $\Sp$ for simplicity. Eq.~\eqref{eq:LeslieEricksenSI}, on the other hand, is the Leslie-Ericksen equation describing the dynamics of the tangent unit vector field, with
\begin{equation}
h_{\parallel i} 
= -\frac{\delta F}{\delta p_{\parallel}^{i}} 
= \kF\nabla^{2}p_{\parallel i}\;,
\end{equation}
the covariant molecular field and $u_{\parallel ij} = (\nabla_{i} v_{\parallel j}+\nabla_{j} v_{\parallel i})/2$ and $\omega_{\parallel ij} = (\nabla_{i} v_{\parallel j}-\nabla_{j} v_{\parallel i})/2$ the covariant strain rate and vorticity tensor respectively. Finally, force balance along the surface normal direction, leads to the shape equation Eq.~\eqref{eq:shape}, where
\begin{multline}
f_{\rm e}^n = 2 \gamma H \\- \kB \left\{\Delta H - (H-H_0) \left[2 H (H-H_0) -4H^2 + 2K\right]\right\}
\end{multline}
is the normal force per unit area arising in response of a departure from the minimizer of the Helfrich energy and
\begin{equation}
f_{\rm p}^n = 2 \kF (2Hg^{ij}-K^{ij})\nabla_i\nabla_j\chi + 2 \kF (K^{ij}-Hg^{ij})\nabla_i\chi\nabla_j\chi\;,
\end{equation}
the normal force per unit area resulting from a distortion of the director $\bm{p}_\parallel$. The function $\chi$ is known as geometric potential and can be found from the Poisson equation $\nabla^2 \chi = K - \rho_{\rm d}$, where $\rho_{\rm d}$ is the topological charge density~\cite{Bowick2009}.

\subsection{\label{sec:Appendix_SphereDeformation} Polar spherical shells}

A sphere of radius $R_0$ is parametrized in spherical coordinates as
\begin{equation}
\bm{X}= R_0 \begin{pmatrix} \sin \theta \cos \varphi \\ \sin \theta \sin \varphi \\ \cos \theta \end{pmatrix}
\end{equation}
with $0 \le \theta \le \pi$ and $0 \le \varphi < 2\pi$. A stationary configuration of the polarization field $\bm{p}_{\parallel}$ can then be constructed in the form
\begin{equation}\label{eq:p_parallel}
\bm{p}_{\parallel} = \frac{1}{R_0}\left(\cos\epsilon\,\bm{e}_{\theta}+\frac{\sin\epsilon}{\sin\theta}\,\bm{e}_{\varphi}\right)\;,
\end{equation}
where $\epsilon$ is a constant determining the local geometry of the two $+1$ defects located at the poles of the sphere. If $\epsilon = 0$ the polarization field is along the meridians of the sphere and the defects are asters. On the other hand, if $\epsilon = \pi/2$ the polarization field is purely azimuthal and the defects are vertices.

With Eq.~\eqref{eq:p_parallel} in hand, one can solve Eqs.~\eqref{eq:NavierStokesSI} and \eqref{eq:LeslieEricksenSI} to find the lateral pressure $\Pi$ and velocity $\bm{v}_{\parallel}$. Taking the divergence of Eq.~\eqref{eq:NavierStokesSI} gives
\begin{equation}
\nabla^{2} \Pi = \alpha \nabla_i \nabla_j p^i_{\parallel} p^j_{\parallel}\;,
\end{equation}
whose azimuthally symmetric solution, i.e. $\Pi=\Pi(\theta)$, is found in the form
\begin{equation}
\label{eq:PressureFixedSphereSI}
\Pi = P_0 + P_1 \artanh(\cos\theta) + \alpha \cos(2\epsilon) \log(\sin\theta)\;,
\end{equation}
with $P_0$ and $P_1$ integration constants. Similarly, using that an azimuthally symmetric incompressible velocity field must vanish along the $\theta-$direction, we set $\bm{v}_{\parallel}=v^{\varphi}\bm{e}_{\varphi}$, with $v^{\varphi}=v^{\varphi}(\theta)$. Thus, from the $\theta-$component of Eq.~\eqref{eq:NavierStokesSI} we then find $P_1 = 0$ and
\begin{equation}
\eta \tan\theta\,\partial_\theta^2 v^\varphi_{\parallel} + 3 \eta\,\partial_\theta v^\varphi_{\parallel}  + \alpha\,\frac{\sin 2\epsilon}{\sin \theta} = 0 \;,
\end{equation}
which has the solution
\begin{multline}
v^\varphi_{\parallel} = \frac{c_2}{2} \left[\artanh(\cos\theta)+\frac{\cot\theta}{\sin\theta}\right] \\ 
+ \frac{\alpha \cos \epsilon \sin \epsilon}{2\eta} \left[\artanh(\cos\theta)-\frac{\cot\theta}{\sin\theta}\right] \;.
\end{multline}
Setting $c_2 = \alpha \cos \epsilon \sin\epsilon/\eta$ to cancel the divergent term we thus arrive at
\begin{equation}
\label{eq:VelocityFixedSphereSI}
v^\varphi_{\parallel} = \frac{\alpha \sin 2\epsilon}{2\eta}\,\artanh(\cos\theta) \;.
\end{equation}
To find the physical velocity field in the coordinate system of $\mathbb{R}^3$ we have to multiply this velocity by $\sin \theta$ and then the velocity is finite everywhere.

Now we turn to the Leslie-Ericksen equation, Eq. \eqref{eq:LeslieEricksenSI}. We find that the equation can be written as $\left(\lambda \cos 2\epsilon + 1 \right) \partial_\theta v^\varphi_{\parallel} = 0$ and therefore
\begin{equation}
\epsilon = \frac{1}{2}\arccos\left(-\frac{1}{\lambda}\right)\;,
\end{equation}
so that we find that a stationary solution is possible only if $|\lambda| > 1$ and that the geometry of the director field is set by the flow alignment parameter. It is instructive to replace these solutions into the shape equation Eq.~\eqref{eq:shape} to compute the elastic pressure necessary to keep the sphere from deforming if that was allowed. We find that away from the poles
\begin{equation}
\Delta P = \frac{2}{R_0} \left[\gamma + \frac{\alpha}{\lambda} \log\left(\sin \theta\right) + \frac{\kF}{R_0^2}\right]\;,
\end{equation}
which shows the different scaling with $R_0$ that can also be found from dimensional analysis. Thus surface tension and activity  dominate at large $R_{0}$ values, whereas at small $R_{0}$ values Frank elasticity is dominant.

Before moving on to the deformable sphere we note that the pressure field of an isolated $+1$ defect in a flat disk of radius $R$ is given by $\Pi = - \alpha/\lambda \log(r/R)$ (see e.g. Ref.~\cite{Kruse2004}) where $r$ is the radial distance from the defect center. This is equal to the pressure near the center of a $+1$ defect at the poles of a sphere. Namely, up to additive constants, $\Pi = - \alpha/\lambda \log(\sin\theta) \simeq - \alpha/\lambda \log \theta + \mathcal{O}\left(\log^2\theta\right)$ where $\theta$ can be seen (to first order) as the radial distance from the defect.

\subsection{Flattening of polar spherical shells}

To assess the stability of active shells with respect to flattening, we express the normal force per unit area $f_{\rm p}^{n}$, featured in Eq.~\eqref{eq:shape}, in terms of Legendre polynomials $\mathcal{P}_l=\mathcal{P}_l(\theta)$. That is
\begin{equation}
f_{\rm p}^n = - \frac{1}{R_0^3} \sum_{l>0} \frac{(l-1)(l+2)}{l(l+1)}\,\mathcal{P}_l s_l \;,
\end{equation}
where $s_l = \sum_i \mathcal{P}^*_l (\theta_i)$ is the defect topological charge density of an aster or vortex (see e.g. Refs. \cite{Lenz2003,GarciaAguilar_2020}).
\begin{figure*}
    \centering
    \includegraphics[width=\textwidth]{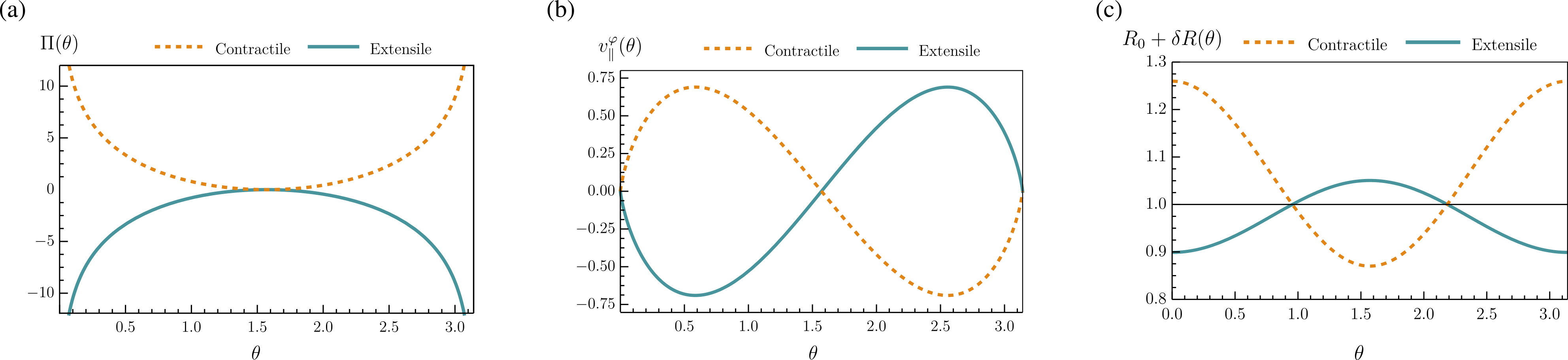}
    \caption{In panels (a) and (b) we show, respectively, the pressure $\Pi$ (Eq. \eqref{eq:PressureFixedSphereSI}) and physical velocity field $v^\varphi_{\parallel}(\theta)$ (Eq. \eqref{eq:VelocityFixedSphereSI} multiplied by $\sin \theta$) for a fixed sphere for contractile and extensile activity. In panel (c) we show the radius of the deformed sphere for extensile and contractile activity, see Eq. \eqref{eq:DeformationSphereSI}. The black horizontal line corresponds to $\delta R(\theta) = 0$. If the extensile activity is sufficiently large, as it is here, the sphere flattens at the poles, that is the radius after deformation is less than $R_0$. The parameter values used for all panels are $\alpha = \pm 5$, $\lambda = 1.1.$, $\eta = 1$, $\kF = 1$, $\gamma=1$, $\kB = 1$, $R_0=1$.}
    \label{fig:SIfig2}
\end{figure*}
We next look at radial perturbations $R = R_0 + \delta R(\theta)$ with respect to the initial shape. This gives the following expressions for the mean and Gaussian curvature at the linear order in $\delta R$:
\begin{subequations}
\begin{gather}
H = \frac{R_0 - \delta R}{R_0^2} - \frac{\nabla^{2} \delta R}{2}\;,\\
K = \frac{R_0 - 2 \delta R}{R_0^3} - \frac{\nabla^{2} \delta R}{R_0}\;, 
\end{gather}
\end{subequations}
where $\nabla^{2}=1/(R_{0}^{2}\sin\theta)\,\partial_{\theta}\sin\theta\,\partial_{\theta}+1/(R_{0}\sin\theta)^{2}\,\partial_{\varphi}^{2}$ is the unperturbed covariant Laplacian. Analogously, for the director we find
\begin{equation}
\bm{p}_\parallel = \frac{R_0 - \delta R}{R_0^2} \left(\cos\epsilon\,\bm{e}_{\theta}+\frac{\sin\epsilon}{\sin\theta}\,\bm{e}_{\varphi}\right)
\end{equation}
such that it is normalized with respect to the perturbed metric. For the lateral pressure we then find instead
\begin{align}
\Pi = P_0 + P_1 \artanh(\cos \theta) + \alpha \cos 2\epsilon \left[\log(\sin\theta) + \frac{\delta R}{R_0}\right]\;,
\end{align}
and we set $P_0 = P_1 = 0$. 

The equilibrium Helfrich force per unit area is given by
\begin{align}
f_{\rm e}^{n} = 2 \gamma H + \frac{\kB}{2 R_0^4} \left(R_0^4 \nabla^{2} \delta R + 2 \delta R \right) \;.
\end{align}
Next, taking $\delta R = \sum_{l>0} \delta R_l \mathcal{P}_l$ and, using $\nabla^{2}\mathcal{P}_l = -l(l+1)/R_0^2 \mathcal{P}_l$, we find
\begin{multline}
f_{\rm e}^n = 
\frac{2 \gamma}{R_0} \\
+ \sum_{l>0} \frac{\delta R_l \mathcal{P}_l}{R_0^2} \left\{\gamma (l-1)(l+2) 
+ \frac{\kB}{R_0^2} \left[l^2(l+1)^2 + 2\right]\right\} \;.
\end{multline}
Furthermore, taking $\Pi H \approx \Pi/R_0$ for small $\alpha$ values, one can recast Eq.~\eqref{eq:shape} in the form
\begin{align}
\Delta P 
&= \frac{2 \gamma}{R_0}+\frac{2 \alpha}{\lambda R_0} \log(\sin\theta) \notag \\
&+ \sum_{l>0} \frac{\delta R_l \mathcal{P}_l}{R_0^2} \left\{\gamma (l-1)(l+2) \vphantom{\frac{\kB}{R_0^2}} 
+ \frac{\kB}{R_0^2} \left[l^2(l+1)^2 + 2\right]\right\} \notag \\
&- \frac{\kF}{R_0^3} \sum_{l>0} \frac{(l-1)(l+2)}{l(l+1)}\,\mathcal{P}_l s_l \;.
\end{align}
Finally, using the expansion
\begin{equation}
\log(\sin \theta) = - \frac{1}{\sqrt{4 \pi}}-\sum_{l \in 2\mathbb{N}} \frac{2\sqrt{(2l+1)\pi}}{l(l+1)}\,\mathcal{P}_l\;, 
\end{equation}
where $2\mathbb{N}=\{2,\,4,\,6\,\ldots\}$ denotes the set of even natural numbers. From the zero mode, on the other hand, we obtain the renormalized isotropic Laplace pressure
\begin{equation}
\Delta P = \frac{2 \gamma}{R_0} - \frac{\alpha}{\lambda R_0 \sqrt{\pi}} \;.
\end{equation}
Now, for a $+1$ defect at each pole one has
\[
s_l = 2\pi \left[\mathcal{P}_l(0) + \mathcal{P}_l(\pi)\right]\;,
\]
hence $s_l = 0$ if $l$ is odd and $s_l = \sqrt{4\pi(2l+1)}$ if $l$ is even. Thus, for odd $l$ values we find $\delta R_l = 0$ from the shape equation. On the other hand, for even $l$ values we recover Eq.~\eqref{eq:delta_R}. The dominant mode is $l=2$, thus, neglecting higher modes, yields
\begin{equation}
\label{eq:DeformationSphereSI}
\delta R_{2} = \frac{5}{4}\left(\frac{2 \kF R_0 + \frac{\alpha}{\lambda} R_0^3}{6 R_0 \gamma + 57 \kB}\right)\left(3 \cos^2\theta - 1\right) \;.
\end{equation}
Hence, when $R_{0}>R_{\rm c}$, with $R_{\rm c}$ given in Eq.~\eqref{eq:critical_radius}, the radial displacement at the poles, i.e. $\theta=0,\,\pi$, changes from positive to negative, thus marking the flattening transition of the active shell.

\bibliography{main.bib} 

\end{document}


\title{Tuneable defect-curvature coupling and topological transitions in active shells \\ Supplementary Information}
\author{Ludwig A. Hoffmann$^1$}
\author{Livio Nicola Carenza$^1$}
\author{Luca Giomi$^1$}
\email{giomi@lorentz.leidenuniv.nl}
\affiliation{$^1$ Instituut-Lorentz, Universiteit Leiden, P.O. Box 9506, 2300 RA Leiden, The Netherlands}
\date{\today}
\maketitle

\section{Movies}

\textbf{Movie 1: Flattening of polar shell in presence of extensile activity.} The extensile active stresses ($\alpha = -0.002$) lead to a flattening of the initially spherical shell. The two $+1$ defects have a spiral geometry and move away from the poles and the shell becomes motile. The vectors denote the polarization field, while the color code refers to the local magnitude of the polarization according to the color bar at the bottom.

\textbf{Movie 2: Genus transition of polar shell.} For large extensile activity ($\alpha = -0.012$) the a chaotic regime is entered and the shell is deformed significantly. Small vesicles separate from the original shell and flatten, eventually leading to a genus transition from a spherical topology to a toroidal topology.

\textbf{Movie 3: Flattening of polar shell in presence of contractile activity.} The contractile active stresses ($\alpha = 0.004$) lead to a flattening of the initially spherical shell albeit less than in the presence of comparable extensile stresses. The two $+1$ defects have an aster geometry and near the center of the defects we observe a buckling of the shell. The defects move away from the poles and the shell becomes motile.

\textbf{Movie 4: Periodic deformation of nematic shell.} In the case of small activity ($\alpha = -0.002$ in this movie) the periodic movement of the four active, motile $+1/2$ defects couples to the elastic shell resulting in a periodic deformation of the shell.

\textbf{Movie 5: Tentacle formation.} In the presence of intermediate extensile activity ($\alpha = -0.004$ in this movie) we observe the periodic deformation of the shell as well as the creation of tentacles which are formed by two $+1/2$ defects approaching each other and in the process creating a protrusion.

\textbf{Movie 6: Chaotic deformation of nematic shell for extensile activity.} For large extensile activity ($\alpha = -0.007$ in this movie) the original shell is elongated and eventually the shell rips apart creating several smaller snail-like surfaces that eventually dissolve due to Oswald ripening.

\textbf{Movie 7: Spindle-like shape of nematic shell in presence of contractile activity.} For intermediate contractile activity two $+1/2$ defects move towards the poles and the sphere is deformed into a spindle-like shape.